\documentclass{ws-procs9x6}

\begin{document}

\title{Instanton model of light hadrons: from low to high energies \thanks{%
\uppercase{T}alk given at \uppercase{XXXII} \uppercase{I}nternational Simposium Multiparticle Dynamics,
\uppercase{S}eptember 7-13, \uppercase{A}lushta, \uppercase{C}rimea,
\uppercase{U}kraine} \thanks{\uppercase{T}his work is
supported by \uppercase{RFBR G}rants 01-02-16431 and 02-02-16194 and by
\uppercase{G}rant \uppercase{INTAS}-2000-366.}}
\author{A.E. Dorokhov}

\address{
Bogoliubov Laboratory of Theoretical Physics,\\
Joint Institute for Nuclear Research,\\
141980 Dubna, Russia\\
E-mail: dorokhov@thsun1.jinr.ru}
\maketitle

\abstracts{Instanton liquid model being effective model of the QCD vacuum describes well
the hadrons at low as well at intermediate energies. Thus, contact
with perturbative QCD results is possible providing the unique information
about the quark-gluon distribution functions in the QCD vacuum and
hadrons at low energy normalization point.
As an illustrative example we consider the pion transition form factor for the process
$\gamma ^{\star }\gamma^{\star}\rightarrow \pi ^{0}$ at space-like values of photon momenta.
The leading and next-to-leading order power asymptotics of the form factor and
the relation between the light-cone pion distribution
amplitudes of twists 2 and 4 and the dynamically generated quark mass are found.}

The interest to the pion transition form factor has recently revived due to
its measuring by CLEO collaboration\cite{Gronberg:1997fj} at large virtuality of
one of the photons. The knowledge of the form factor at arbitrary photon
virtualities would provide important information about the hadron
contribution to the muon anomalous moment $g-2$ and some other processes.
Theoretically, the pion form factor $M_{\pi ^{0}}(q_{1}^{2},q_{2}^{2})$ for
the transition process $\gamma ^{\star }(q_{1})\gamma ^{\star
}(q_{2})\rightarrow \pi ^{0}(p)$, where $q_{1}$ and $q_{2}$ are photon
momenta, is related to fundamental properties of QCD dynamics at low and
high energies. At zero photon virtualities the value of the form factor and
its slope (radius) is estimated within the chiral perturbative theory. In the
opposite limit of large photon virtualities the leading momentum power
dependence\cite{Lepage:1979zb} of the form factor supplemented by small radiative\cite{delAguila:1981nk}
and power corrections\cite{Chernyak:1982is} is dictated by perturbative QCD (pQCD).

In the following, we discuss
the appoach\cite{Anikin:rq} that allow us to match these extremes and describe the
intermediate energy region. This approach describes quark-meson dynamics
within the effective model, where the quark-quark interaction induced by
instanton exchange leads to spontaneous breaking of the chiral symmetry.
It dynamically generates the momentum dependent quark mass $M(k^{2})$
that may be related to the nonlocal quark condensate\cite{Dorokhov:1997iv}.
Specifically, we find\cite{Dorokhov:2002iu} the pion transition form factor in
wide kinematical region up to moderately large photon virtualities and extract from its
asymptotics the pion distribution amplitudes (DAs) at normalization scale
typical for hadrons.

The invariant amplitude for the process $\gamma ^{\ast }\gamma ^{\ast
}\rightarrow \pi ^{0}$ is given by
\[
A\left( \gamma ^{\ast }\left( q_{1},\epsilon _{1}\right) \gamma ^{\ast
}\left( q_{2},\epsilon _{2}\right) \rightarrow \pi ^{0}\left( p\right)
\right) =-ie^{2}\varepsilon _{\mu \nu \rho \sigma }\epsilon _{1}^{\mu
}\epsilon _{2}^{\nu }q_{1}^{\rho }q_{2}^{\sigma }M_{\pi ^{0}}\left(
q_{1}^{2},q_{2}^{2}\right) ,
\]%
where $\epsilon _{i}^{\mu }(i=1,2)$ are the photon polarization vectors.
Consider first the low energy region.
With both photons real $\left( q_{i}^{2}=0\right) $ one finds the result
\begin{equation}
M_{\pi ^{0}}\left( 0,0\right) =\frac{N_{c}}{6\pi ^{2}f_{\pi }}%
\int_{0}^{\infty }du\frac{uM(u)\left[ M(u)-2uM^{\prime }(u)\right] }{D^{3}(u)
}=\frac{1}{4\pi ^{2}f_{\pi }},  \label{ChAn}
\end{equation}
where $D(u)=u+M^{2}(u)$ and $M^{\prime }(u)=\frac{d}{du}M(u),$
consistent with the chiral anomaly and independent of the shape of $M(k^{2}).$
Below, for the numerical analysis we choose the dynamical quark mass profile in the Gaussian
form $M_{G}(k^{2})=M_{q}\exp {(-2k^{2}/\Lambda ^{2})},$ where we take
$M_{q}=350$ MeV and fix $\Lambda =1.29$ GeV from the pion weak decay constant,
$f_{\pi }=92.4$ \textrm{MeV}. We also consider the shape given by the quark
zero modes (z.m.) in the instanton field: $M_{I}(k^{2})=M_{q}Z^{2}(k\rho )$,
where $Z(k\rho )=2z\left[
I_{0}(z)K_{1}(z)-I_{1}(z)K_{0}(z)-I_{1}(z)K_{1}(z)/z\right] _{z=k\rho /2}$, with
$\rho =1.7$ GeV$^{-1}$ being the inverse mean instanton radius and $M_{q}=345$ MeV.
The mean square radius of the pion for the transition $\gamma ^{\ast }\pi
^{0}\rightarrow \gamma $ is found to be $r_{\pi \gamma }^{2}\approx 1/(2\pi^2f_\pi^2)$
and is almost independent on the form of $M(k^{2}).$

At large photon virtualities $Q^{2}=-(q_{1}^{2}+q_{2}^{2})$ the model calculations reproduce the pQCD factorization
result ($\omega=(q_{1}^{2}-q_{2}^{2})/(q_{1}^{2}+q_{2}^{2})$)
\begin{equation}
\left. M_{\pi ^{0}}(q_{1}^{2},q_{2}^{2})\right\vert _{Q^{2}\rightarrow
\infty }=J^{(2)}\left( \omega \right) \frac{1}{Q^{2}}+J^{(4)}\left( \omega
\right) \frac{1}{Q^{4}}+O(\frac{\alpha _{s}}{\pi })+O(\frac{1}{Q^{6}}).
\label{AmplAsympt}
\end{equation}%
The leading and next-to-leading order asymptotic
coefficients
\begin{eqnarray}
J^{(2)}\left( \omega \right)  &=&\frac{4}{3}f_{\pi }\int_{0}^{1}dx\frac{%
\varphi _{\pi }^{(2)}(x)}{1-\omega ^{2}(2x-1)^{2}},\ \ \   \nonumber \\
J^{(4)}\left( \omega \right)  &=&\frac{4}{3}f_{\pi }\Delta ^{2}\int_{0}^{1}dx%
\frac{1+\omega ^{2}(2x-1)^{2}]}{[1-\omega ^{2}(2x-1)^{2}]^{2}}\varphi _{\pi
}^{(4)}(x)  \label{J}
\end{eqnarray}%
are expressed in terms of the light-cone pion distribution amplitudes (DA),
$\varphi _{\pi }(x)$, that are predicted by the model at the low normalization
scale $\mu ^{2}\sim \Lambda ^{2}\sim \rho^{-2}$ (Fig. 1, for Gaussian $M(k^2)$)
\[\varphi _{\pi }^{(2)}(x)=\frac{N_{c}}{4\pi ^{2}f_{\pi }^{2}}\int_{-\infty
}^{\infty }\frac{d\lambda }{2\pi }\int_{0}^{\infty }du\cdot
\]
\begin{equation}
\cdot\frac{F(u+i\lambda
\overline{x},u-i\lambda x)}{D\left( u-i\lambda x\right) D\left( u+i\lambda
\overline{x}\right) }\left[ xM\left( u+i\lambda \overline{x}\right) +\left(
x\leftrightarrow \overline{x}\right) \right] ,  \label{WF_VF2}
\end{equation}
\[\varphi _{\pi }^{(4)}(x)=\frac{1}{\Delta ^{2}}\frac{N_{c}}{4\pi ^{2}f_{\pi
}^{2}}\int_{-\infty }^{\infty }\frac{d\lambda }{2\pi }\int_{0}^{\infty }du\cdot
\]
\begin{equation}
\cdot\frac{uF(u+i\lambda \overline{x},u-i\lambda x)}{D\left( u-i\lambda x\right)
D\left( u+i\lambda \overline{x}\right) }\left[ \overline{x}M\left(
u+i\lambda \overline{x}\right) +\left( x\leftrightarrow \overline{x}\right) %
\right] ,  \label{WF_VF4}
\end{equation}
where
$F\left( u,v\right) =\sqrt{M\left( u\right) M\left( v\right) }.$
The parameter $\Delta ^{2}$  characterizing the scale of the power corrections
in the hard exclusive processes is
\begin{equation}
\Delta ^{2}=\frac{N_{c}}{4\pi ^{2}f_{\pi }^{2}}\int_{0}^{\infty }du\frac{%
u^{2}M(u)(M(u)+\frac{1}{3}uM^{\prime }(u))}{D^{2}(u)},  \label{PowCorr}
\end{equation}
Its value is predicted $\Delta^{2}= 2.41(2.74)\pi^2f_\pi^2$  for Gaussian
(zero mode) shape of $M(k^2)$, correspondingly.
As it is clear from Fig. 1, the leading order
pion DA, $\varphi _{\pi }^{(2)}(x)$, is close to the asymptotic form that is in
agreement with the results obtained previously in\cite{Mikhailiov:1989mk,Kroll:1996jx}.
In the leading order the similar results within the instanton model have been
derived earlier in\cite{Esaibegian:1989uj}.

The asymptotic coefficients $J(\omega )$  may be written in the form
\[J^{(2)}\left( \omega \right) =-\frac{1}{\pi^2f_\pi}\int_{0}^{\infty
}duu\int_{0}^{\infty }dv\cdot
\]
\begin{equation}
\cdot\left\{ \frac{M^{1/2}\left( z_{-}\right) }{D(z_{-})}%
\frac{\partial }{\partial z_{+}}\left( \frac{M^{3/2}\left( z_{+}\right) }{%
D(z_{+})}\right) +\left( z_{-}\longleftrightarrow z_{+}\right) \right\} ,
\label{J2}
\end{equation}
\[J^{(4)}\left( \omega \right) =\frac{2}{\pi^2f_\pi}\int_{0}^{\infty
}du\int_{0}^{\infty }dvv\cdot
\]
\begin{equation}
\cdot\left\{ \frac{M^{1/2}\left( z_{-}\right) }{D(z_{-})}%
\left[ \frac{M^{3/2}\left( z_{+}\right) }{D(z_{+})}+u\frac{\partial }{%
\partial z_{+}}\left( \frac{M^{3/2}\left( z_{+}\right) }{D(z_{+})}\right) %
\right] +\left( z_{-}\longleftrightarrow z_{+}\right) \right\} ,  \label{J4}
\end{equation}%
where $z_{\pm }=u+v(1\pm \omega )$. With the model parameters given above we find
for the process $\gamma \gamma ^{\ast }\rightarrow \pi ^{0}$
the values $J^{(2)}\left( \omega =1\right) =1.83(2.13)f_\pi$
consistent with the CLEO fit $J_{\exp }^{(2)}\left( 1\right) =(1.74\pm 0.32)f_\pi$
and the power correction
$J^{(4)}\left( 1\right) /J^{(2)}\left( 1\right) =2.97(3.62)\pi^{2}f_\pi^2$.

\begin{figure}[tbp]
\vskip -0.5cm \centering
\begin{minipage}[c]{7cm}
\epsfbox{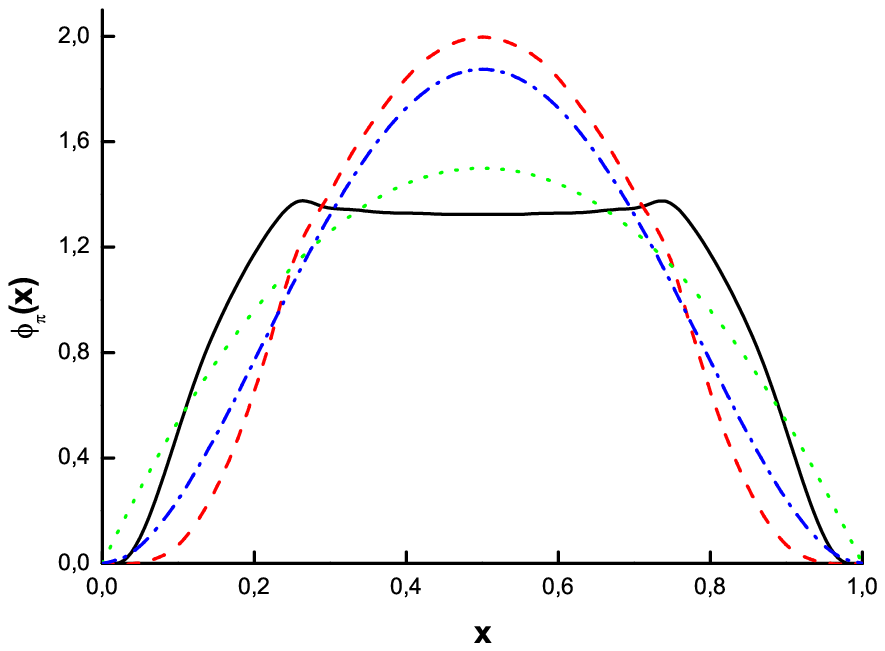}
\caption{  The pion distribution amplitudes (normalized by unity):
the model predictions for twist-2 (solid line) and twist-4 (dashed line) components
and the asymptotic limits of twist-2  (dotted line) and twist-4  (dash-dotted line) amplitudes.
\label{relx}}
\end{minipage}
\hspace*{0.5cm}
\begin{minipage}[c]{7cm}
\epsfbox{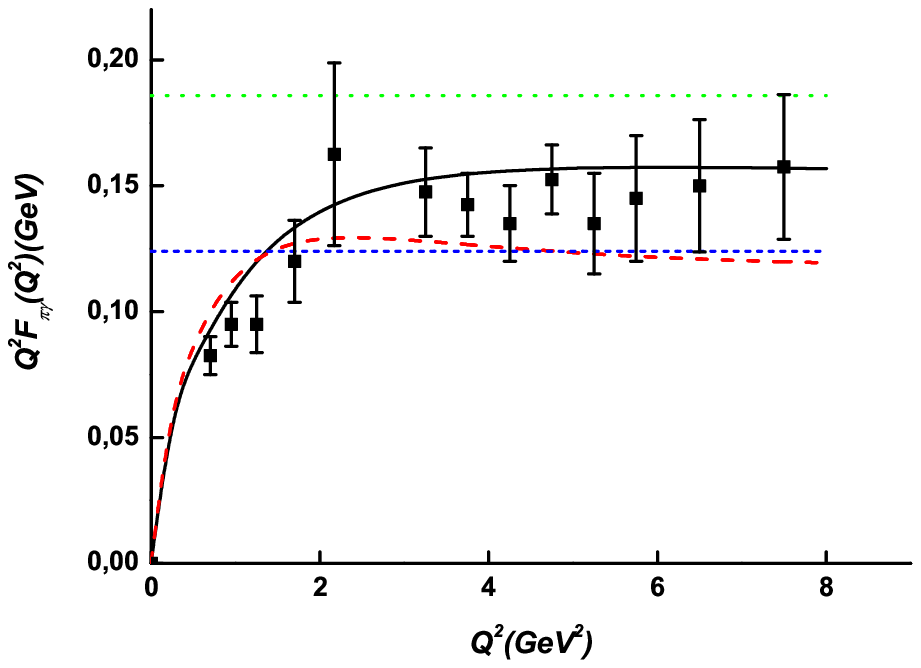}
\caption{ The pion-photon transition form factors $Q^2F_{\pi\gamma^\ast}(Q^2)$
 (solid line) and
$F_{\pi \gamma ^{\ast }\gamma ^{\ast}}(Q^{2})=M_{\pi ^{0}}\left(-Q^{2}/2,-Q^{2}/2\right)$
(dashed line) and their perturbative limits $2f_\pi$ (dotted line) and $4f_\pi/3$ (short dashed line).
The experimental points ($Q^2F_{\pi\gamma^\ast}$) are taken from [1].
\label{rels}}
\end{minipage}
\end{figure}

The model form factors presented in Fig. 2 (for Gaussian $M(k^2)$) take into account the perturbative
$\alpha_{s}(Q^{2})-$ corrections\cite{delAguila:1981nk} to the leading twist-2 term with
the running coupling that has zero at zero momentum. The perturbative corrections to the
twist-4 contribution and the power corrections
generated by the twist-3 pion DAs are expected to be inessential.

In conclusion, within the covariant nonlocal model describing the quark-pion
dynamics we obtain the $\pi \gamma ^{\ast }\gamma ^{\ast }$ transition form
factor in the region up to moderately high momentum transfer squared, where
the rapid power-like asymptotics takes place. At larger virtualities
the pQCD evolution of the DA slowly goes to the asymptotic limits.
From the comparison of the kinematical dependence of the asymptotic
coefficients of the transition pion form factor, as it is given by pQCD and
the nonperturbative model, the relations
between the pion DAs and the dynamical quark mass and quark-pion vertex are
derived.

\section*{Acknowledgments}

I am grateful to A.M. Bakulev, W. Broniowski, A. Di Giacomo, A.S. Gorski,
N.I. Kochelev, S.V. Mikhailov, M.K. Volkov, L. Tomio and V.L. Yudichev for
many useful discussions on topics related to this talk.


\begin{thebibliography}{0}

\bibitem{Gronberg:1997fj}
J.~Gronberg {\it et al.}  [CLEO Collaboration],
Phys.\ Rev.\ D {\bf 57}, 33 (1998).

\bibitem{Lepage:1979zb}
G.~P.~Lepage and S.~J.~Brodsky,
Phys.\ Lett.\ B {\bf 87}, 359;
Phys.\ Rev.\ D {\bf 22}, 2157 (1980).


\bibitem{delAguila:1981nk}
F.~del Aguila and M.~K.~Chase,
Nucl.\ Phys.\ B {\bf 193}, 517 (1981);
E.~Braaten,
Phys.\ Rev.\ D {\bf 28}, 524 (1983);
E.~P.~Kadantseva, S.~V.~Mikhailov and A.~V.~Radyushkin,
Sov.\ J.\ Nucl.\ Phys.\  {\bf 44}, 326 (1986).

\bibitem{Chernyak:1982is}
V.~L.~Chernyak, A.~R.~Zhitnitsky and I.~R.~Zhitnitsky,
Sov.\ J.\ Nucl.\ Phys.\  {\bf 38}, 645 (1983);
V.~A.~Novikov, M.~A.~Shifman, A.~I.~Vainshtein, M.~B.~Voloshin and V.~I.~Zakharov,
Nucl.\ Phys.\ B {\bf 237}, 525 (1984);
A.~S.~Gorsky,
Sov.\ J.\ Nucl.\ Phys.\  {\bf 50}, 498 (1989).

\bibitem{Anikin:rq}
See for a review, \textit{e.g.}, I.~V.~Anikin, A.~E.~Dorokhov and L.~Tomio,
Phys.\ Part.\ Nucl.\  {\bf 31}, 509 (2000).

\bibitem{Dorokhov:2002iu}
A.~E.~Dorokhov,
arXiv:hep-ph/0212156.

\bibitem{Dorokhov:1997iv}
A.~E.~Dorokhov, S.~V.~Esaibegian and S.~V.~Mikhailov,
Phys.\ Rev.\ D {\bf 56}, 4062 (1997);
A.~E.~Dorokhov and W.~Broniowski,
Phys.\ Rev.\ D {\bf 65}, 094007 (2002).

\bibitem{Mikhailiov:1989mk}
S.~V.~Mikhailiov and A.~V.~Radyushkin,
Sov.\ J.\ Nucl.\ Phys.\  {\bf 52}, 697 (1990);
A.~P.~Bakulev, S.~V.~Mikhailov and N.~G.~Stefanis,
Phys.\ Lett.\ B {\bf 508}, 279 (2001).

\bibitem{Kroll:1996jx}
P.~Kroll and M.~Raulfs,
Phys.\ Lett.\ B {\bf 387}, 848 (1996);
I.~V.~Anikin, A.~E.~Dorokhov and L.~Tomio,
Phys.\ Lett.\ B {\bf 475}, 361 (2000);
M.~Diehl, P.~Kroll and C.~Vogt,
Eur.\ Phys.\ J.\ C {\bf 22}, 439 (2001).

\bibitem{Esaibegian:1989uj}
S.~V.~Esaibegian and S.~N.~Tamarian,
Sov.\ J.\ Nucl.\ Phys.\  {\bf 51}, 310 (1990);
A.~E.~Dorokhov,
Nuovo Cim.\ A {\bf 109}, 391 (1996).

\end{thebibliography}
\end{document}